# Extracting Major Topics of COVID-19 Related Tweets

Faezeh Azizi
Perlab, Faculty of Electrical and Computer Engineering
University of Birjand
Birjand, Iran
faezeh.azizi1995@birjand.ac.ir

Hamed Vahdat-Nejad*
Perlab, Faculty of Electrical and Computer Engineering
University of Birjand
Birjand, Iran
vahdatnejad@birjand.ac.ir

Hamideh Hajiabadi
Department of Computer Engineering
Birjand University of Technology
Birjand, Iran
hajiabadi@birjandut.ac.ir

Mohammad Hossein Khosravi
Faculty of Electrical and Computer Engineering
University of Birjand
Birjand, Iran
mohokhosravi@birjand.ac.ir

*Abstract*—With the outbreak of the Covid-19 virus, the activity of users on Twitter has significantly increased. Some studies have investigated the hot topics of tweets in this period; however, little attention has been paid to presenting and analyzing the spatial and temporal trends of Covid-19 topics. In this study, we use the topic modeling method to extract global topics during the nationwide quarantine periods (March 23 to June 23, 2020) on Covid-19 tweets. We implement the Latent Dirichlet Allocation (LDA) algorithm to extract the topics and then name them with the "reopening", "death cases", "telecommuting", "protests", "anger expression", "masking", "medication", "social distance", "second wave", and "peak of the disease" titles. We additionally analyze temporal trends of the topics for the whole world and four countries. By analyzing the graphs, fascinating results are obtained from altering users' focus on topics over time.

## I. Introduction

Covid-19 is the most critical issue that has severely affected human life in 2020 and 2021. During this pandemic, users' activities on microblogging platforms such as Twitter are remarkable. From January to May 2020, about 628 million tweets, with Coronavirus and Covid19 hashtags, were reported[1]. In general, considerable textual data, such as user tweets, contain valuable information and knowledge. By processing them, pertinent information will be obtainable for many decision-makers and those dealing with Covid-19 and related issues.

One of the most critical procedures for processing and gaining knowledge from Corona-related tweets is obtaining the chief topics or topic modeling [1]. By topic modeling, leading and dominant topics can be extracted from user's tweets. Via the topic modeling, goals like better disease monitoring [2] and measuring the effectiveness of vaccines [3] have been achieved. During the Covid-19 pandemic, some studies conducted topic modeling on Twitter posts and extracted items such as "expression of loneliness" [4], users' attitudes about domestic violence [5], the impact of weather on the spread of the virus [6], the impact of the virus on stock market [7], and users' views on the use of web-conferencing systems [8]. In general, analyzing tweets related to Covid-19 and extracting various topics discussed around the world is essential. In this regard, some studies have been conducted to extract the leading topics from the tweets associated with this virus [9-12].

In most previous studies, changes in the importance of Covid-19 related topics over time have not been sufficiently addressed. Indeed, the Covid-19 discussed topics contain interesting ascents and descents for different countries, which can be useful for decision-makers. We extract the significant topics from a vast number of Covid-19 related tweets, tweeted from March 23 to June 23, 2020, in four countries with the most considerable number of Covid-19 cases in the world[2], namely, the United States, China, United Kingdom, and Canada. We implement the LDA algorithm [13], to extract the top ten topics; namely, "reopening", "death cases", "telecommuting", "protests", "anger expression", "peak of the disease", "masking", "medication", "social distance", and "second wave". Ultimately, we investigate temporal trends of topics separately for the four mentioned countries. These graphs show that in most countries, topics of "reopening" and "death cases" are highlighted, and there is a correlation between them over time.

---

[1] https://www.tweetbinder.com/blog/covid-19-coronavirus-twitter

[2] https://coronavirus.jhu.edu/map.html

The remaining sections of the article are as follows: The second section describes the proposed method. The third section includes the experiment. The fourth section concludes the paper.

## II. PROPOSED METHOD

The proposed method consists of three significant steps, described below:

### A. Compiling tweets

Statistics show that the most used Twitter hashtag in 2020 is dedicated to #Covid19[3]. More than two million Covid-19 associated tweets from March 23 to June 23, 2020, are considered. These 14 weeks coincide with the date of the announcement of the global quarantine law by many countries. Afterward, we parse the tweets into a collection of tokens through the Tokenization process. Next, we remove stop words (like a, about, all, am, etc.), the hashtag sign, user ID, and links from the tokens. In addition, we remove components such as Covid-19, Covid, corona, virus, epidemic, Trump, which maintain a high frequency in all tweets and do not have information for modeling. Finally, we apply stemming and lemmatization to the tokens to convert variant forms of a word (e.g., buying, buying, bought) into a single form (e.g., buy). It also helps to reduce the dictionary size and increases the recall rate for the next step.

### B. Topic modeling

Topic modeling is an unsupervised machine learning technique that detects the word and phrase patterns (e.g., phrases that repeatedly appear simultaneously) in a document (or a set of documents), followed by clustering similar expressions that best describe the dominant topic of that document. Finally, the names of the topics for the clusters are labeled by the researchers (guessing the relationship within a cluster of expressions). Topic modeling algorithms work to furnish bolder topics across the entire document set (rather than each document alone). We use the LDA, as one of the most common algorithms of topic modeling, to extract the key topics of the tweets. We use the common Term Frequency–Inverse Document Frequency (TF-IDF) [15] method for extracting a vector containing the weight of the tokens from all tweets (our training data). Each cluster or set of tokens usually points to a specific topic that can be named. After trial and error on different values (5, 6 … 20), we assume the number of topics to be ten. To get better and more accurate results, we remove the retweets and consider a certain number (150,000) of tweets randomly for each week, so that the weekly comparison of tweets is possible. Table. 1 shows the topics' titles with the most relevant words and an example of users' tweets for each country.

TABLE I. EXAMPLES OF TOPICS

| Topics | The most relevant words | Tweet example |
|---|---|---|
| Telecommuting | work home lockdown stand | " Work home experience summarised in a picture. Stay home! Stay Safe." |
| Death cases | case death report total number | "A other death toll due to corona today... I'm scared " |
| Protests | protest racist black leader right | " India: Muslim protester refuses to disperse, says "corona means Qur'an" and won't harm Muslims" |
| Second wave | second surg warn prepar | "The second wave of Corona is hitting Asia. The second round of bear #equity markets are about to hit as well." |
| Reopening | reopen safe rule increase recoveri | " Gyms should reopen tf. I'll clock in and work during the corona crisis cus' i miss my workouts. " |
| Anger expression | rally shit stupid sick | " i'm sorry.. but y'all are so damn stupid if you're out still PARTYING with this corona virus going around. " |
| Masking | mask hand face | " Kindly use face masks to protect yourself and let's all fight against coronavirus #facemask. " |
| Peak of the disease | spike scientist | "#RESIST Corona virus is just beginning to spike in US, we must do all we.. " |
| Medication | drug dexamethasone save | " Dexamethasone Is Proven to be the 1st Covid-19 cure.... It's not effective in early stages but very " |
| Social distance | social distanc keep | " Keep in mind that social distancing will reduce the need for hospital beds for all other infectious diseases. This…" |

The most relevant words on each topic with an example of the tweet.

### C. Location tagging

The content analysis of users' tweets to extract the location content of each tweet is essential. Usual methods such as the user's Geolocation are not accurate enough; because the location mentioned in a tweet can be different from the location of the tweet sender. To extract the location mentioned in the tweet, we compile a comprehensive dictionary, which contains more than 2,000 words of country names, state names, city names, and more. We develop this dictionary for four countries: the USA, China, UK, and Canada. As an instance, for the USA, there are about 400 related titles, including the names of all states (Alabama, Florida, etc.), cities (Mesa, Oakland, etc.), abbreviated terms (US, NY, etc.), and also we consider different types of using a place name (United States of America, America, etc.). To find country-related titles, we use the GeoNames[4] geographic database, which contains over 20 million location names. By using a natural language processing (NLP) algorithm, we compare the words of this dictionary with the tweets. Then by setting the normalization rules in this algorithm, we identify tweets relate to each country. For example, if a tweet discusses New York, we assign the tweet to the USA, in the process of normalizing the tweet. For location tagging, we consider the emoji flags of countries, too.

---
[3] https://blog.twitter.com/en_us/topics/insights/2020/spending-2020-together-on-twitter.html

[4] https://www.geonames.org/

III. EXPERIMENT

Python programming environment has been used to execute the LDA model, and the GATE NLP framework [16] has been used to tag the location of tweets. Fig. 1 shows the share of each topic during these 14 weeks; reopening topic contains the most considerable number of tweets (650,669 thousand), then topics of death cases, telecommuting, protests, anger expression, peak of the disease, masking, medication, social distance, and the second wave, respectively were important.

Fig. 2 shows the temporal distribution of topics frequency worldwide from March 23 to June 23; the right axis represents the official statistics of Covid-19, and the left axis indicates the frequency of tweets for each topic. While the number of confirmed cases increases, the topics are divided into two categories of affected and unaffected. Discussions on the topics of death cases, medication, peak of the disease, second wave, reopening, anger expression, and masking have increased with the increase in the number of confirmed cases. The topics of telecommuting, protests, and social distance have been included in the unaffected category.

Fig. 3 shows the temporal distribution of the top four topics for the USA, China, the UK, and Canada. In most of these countries, death cases and reopening topics are considered, and there is a good correlation between these two topics. Also, after comparing the trend of topics with the official daily statistics of Covid-19 cases', it is observed that after the peak in the number of confirmed cases, more attention has been paid to the death cases topic. So in the case of the USA, Canada, and the UK, discussion over this topic has peaked. In the USA, the topics of reopening, death cases, medication, and peak of the disease have been in the affected category. In the case of China, the topic of protests has been in the unaffected category, and the reopening, death cases, and peak of the disease topics have been among the affected category; the protests topic against China has always been important, and Twitter users seem to be discussing the country with protesting tweets. In the case of the UK, the topics of reopening, medication, masking, and death cases are among the affected category. For Canada, the second wave, death cases, anger expression, and reopening topics are among the affected category.

IV. CONCLUSION

In this paper, a topic modeling method with tweets content analysis has been presented. With the help of the LDA algorithm, key topics have been extracted from a dataset including more than 2 million tweets related to Covid-19. A dictionary-based location extraction method has been used for geotagging tweets. Topics' temporal diagram of worldwide and four countries have been presented. After analyzing the trend of topics and the trend of Covid-19 confirmed cases in the world, it was discovered that with the increase in the number of confirmed cases, the discussion about some topics increased.

One of the limitations of this work is that it only considers English language tweets. Besides, considering only the experience of Twitter users cannot reflect the opinions of all people in a community. In future research, other languages like Persian, Chinese, Arabic, etc. can be investigated. It is also possible to evaluate people's viewpoints by focusing on specific domains like economics and education.

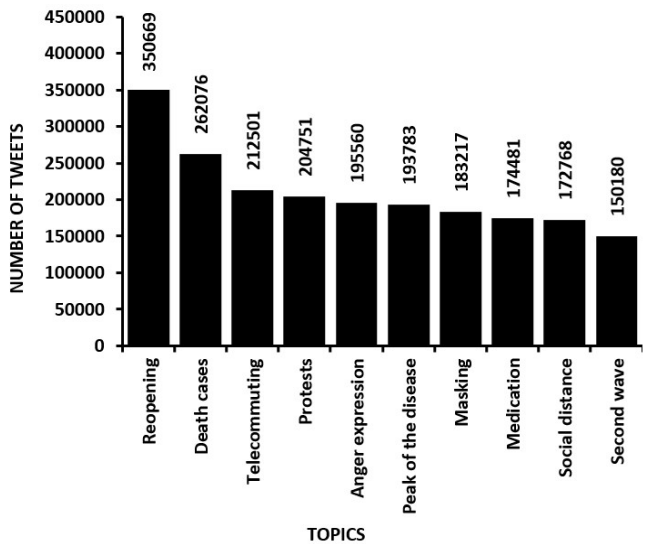

Fig. 1. Number of Tweets of each topic in all tweets.

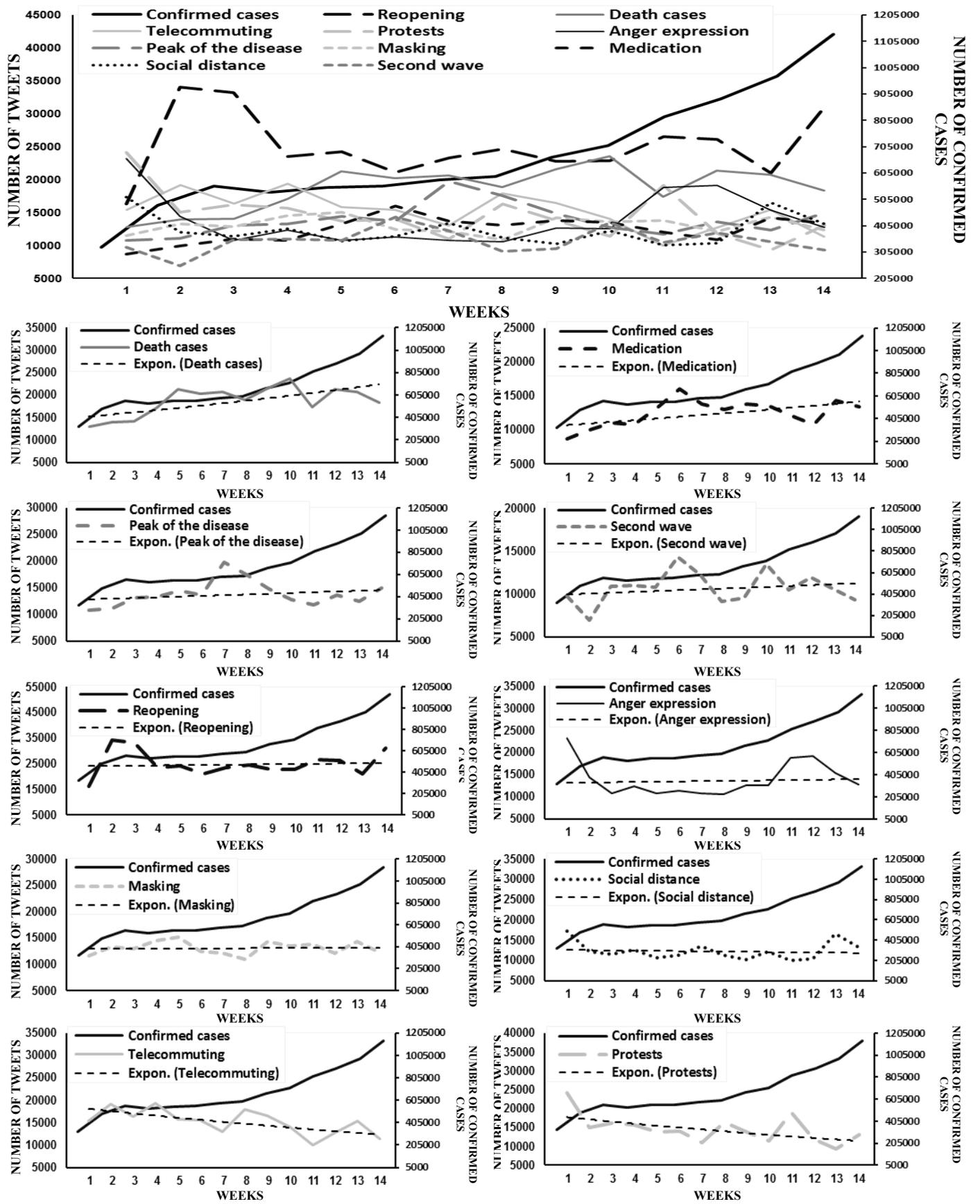

Fig. 2. Temporal distribution of topics

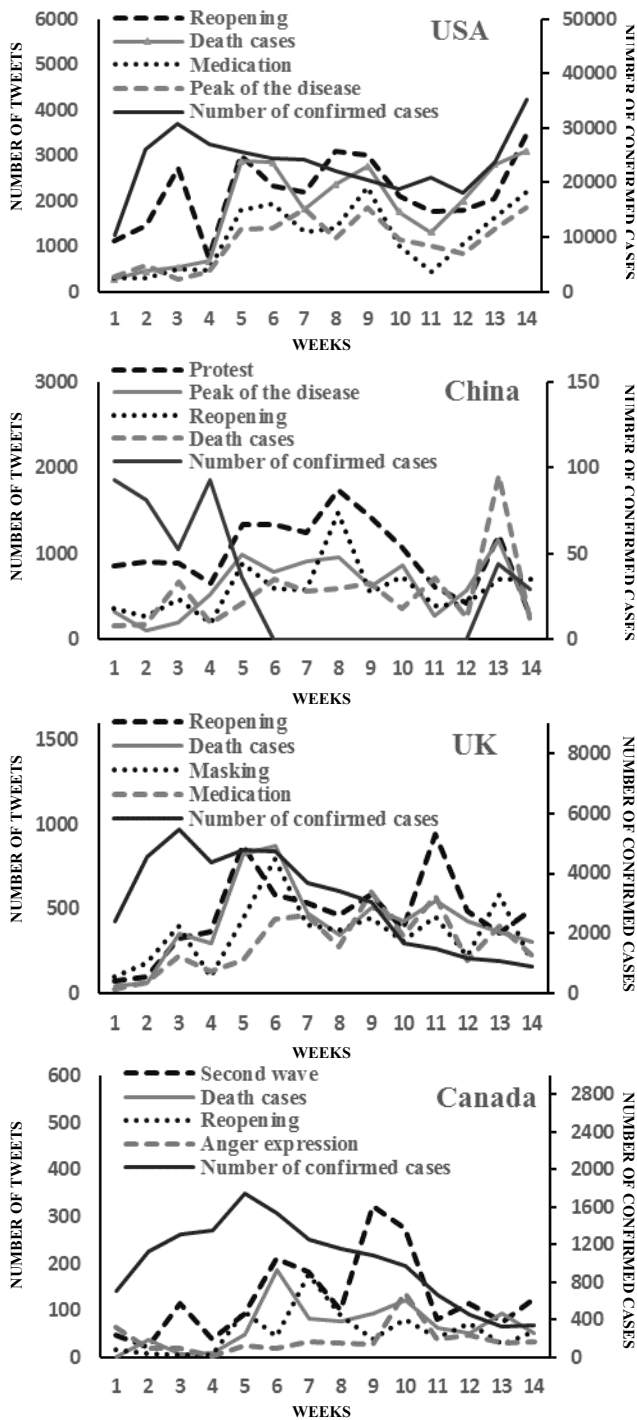

Fig. 3. Temporal distribution of top four topics for 4 countries.

ACKNOWLEDGMENT

It is our pleasure to thank Mr. Parsa Bagherzadeh (currently a Ph.D. candidate at the Concordia University) for his help in the implementation.